\documentclass[conference]{IEEEtran}
\IEEEoverridecommandlockouts

\usepackage{amsmath,amssymb,amsfonts}
\usepackage{algorithmic}
\usepackage{graphicx}
\usepackage{textcomp}

\usepackage[style=numeric]{biblatex}
\usepackage{xcolor}
\usepackage{graphicx}
\usepackage{rotating} 
\usepackage{hyperref}
\usepackage{url}
\usepackage{comment}
\usepackage{listings}
\usepackage{caption}
\usepackage{subcaption}
\usepackage{xcolor}
\usepackage{booktabs}
\usepackage{multirow}
\usepackage{tabularx}
\usepackage{adjustbox}

\usepackage{tikz}
\usetikzlibrary{quantikz2}

\definecolor{codegreen}{rgb}{0,0.6,0}
\definecolor{codegray}{rgb}{0.5,0.5,0.5}
\definecolor{codepurple}{rgb}{0.58,0,0.82}
\definecolor{backcolour}{rgb}{0.95,0.95,0.92}
\definecolor{codecyan}{rgb}{0.0,0.2,1.0}

\lstdefinestyle{mystyle}{
    commentstyle=\textcolor{codegreen},
    keywordstyle=\color{codecyan},
    numberstyle=\tiny\color{codegray},
    stringstyle=\color{codepurple},
    basicstyle=\ttfamily\footnotesize,
    breakatwhitespace=false,    
    breaklines=true,    
    captionpos=b,    
    keepspaces=true,    
    numbers=left,    
    numbersep=2pt,  
    firstnumber=auto,
    numberblanklines=false,
    showspaces=false,
    showstringspaces=false,
    showtabs=false,
    tabsize=2
}
\newcommand{\qv}{QV}

\begin{document}

\title{From Bits to Qubits: Challenges in Classical-Quantum Integration}

\author{
\IEEEauthorblockN{Sudhanshu Pravin Kulkarni}
\IEEEauthorblockA{
\textit{San Francisco State University}\\
San Francisco, CA, USA \\
\href{mailto:skulkarni@sfsu.edu}{skulkarni@sfsu.edu}}
\and
\IEEEauthorblockN{Daniel E. Huang}
\IEEEauthorblockA{
\textit{San Francisco State University}\\
San Francisco, CA, USA \\
\href{mailto:danehuang@sfsu.edu}{danehuang@sfsu.edu}}
\and
\IEEEauthorblockN{E. Wes Bethel}
\IEEEauthorblockA{\textit{San Francisco State University} \\
San Francisco, CA, USA \\
\textit{Lawrence Berkeley National Laboratory}\\
Berkeley, CA, USA \\
\href{mailto:ewbethel@sfsu.edu}{ewbethel@sfsu.edu}}
}

\maketitle
\pagestyle{plain}


\begin{abstract}\label{sec:abstract}
While quantum computing holds immense potential for tackling previously intractable problems, its current practicality remains limited. A critical aspect of realizing quantum utility is the ability to efficiently interface with data from the classical world. This research focuses on the crucial phase of quantum encoding, which enables the transformation of classical information into quantum states for processing within quantum systems. We focus on three prominent encoding models: Phase Encoding, Qubit Lattice, and Flexible Representation of Quantum Images (FRQI) for cost and efficiency analysis. The aim of quantifying their different characteristics is to analyze their impact on quantum processing workflows. This comparative analysis offers valuable insights into their limitations and potential to accelerate the development of practical quantum computing solutions.


\end{abstract}
\section{Introduction}\label{sec:introduction}
Classical computers excel in a broad range of tasks, from artificial intelligence to general-purpose problem-solving, thanks to established architectures and algorithms. In contrast, quantum computers promise exponential speedups for specific complex problems, such as Shor’s algorithm for factoring large numbers~\cite{Shors} and Grover’s unstructured search~\cite{Grover}, driving significant theoretical and experimental advances in quantum computing (QC). Over the last decade, QC has evolved from research to industry, making it essential to understand the interplay between classical and quantum computing. However, characterizing the performance of hybrid classical-quantum programs remains challenging due to the complexity and diversity of such systems.

This study examines the cost and performance of various classical-to-quantum data encoding methods in the context of a unary operation within quantum image processing. We analyze trade-offs across three encoding approaches for classical data, including image-based floating-point data, by evaluating multiple performance metrics such as runtime and QC-specific circuit characteristics like complexity and entanglement.

A central contribution of this work is to highlight the challenges of working with classical data in practical QC applications, where performance considerations differ from those in classical computing. While some quantum algorithms, like Shor’s algorithm~\cite{Shors}, demonstrate clear quantum advantage without significant classical data, areas like quantum image processing~\cite{Yan-Venegas:TwentyYears:2024} and quantum machine learning~\cite{Schuld:MLQC:2021} rely heavily on classical data encoding for quantum platforms. This study identifies multi-dimensional trade-offs in different encoding methods without recommending one approach over another, laying the groundwork for further exploration in QC performance with classical data integration.

Sec.~\ref{sec:related} presents background and related work, including a brief introduction to QC topics and notation, code structure, and work in quantum benchmarking. 
Sec.~\ref{sec:dataEnc} describes three quantum encoding models and their implementation as gate-based circuits using IBM's Qiskit SDK~\cite{Qiskit} suitable for execution on simulators~\cite{qiskit_aersimulator} or gate-based QPUs like those at IBMQ~\cite{IBMQ:2024}.
The study methodology in Sec.~\ref{sec:evaluation} is followed by a presentation of results in Sec.~\ref{sec:findings}.

\section{Related Work}\label{sec:related}
This section reviews QC fundamentals and discusses the hybrid classical-quantum computing architecture.
It then presents a detailed discussion of three different quantum encoding techniques: The Phase Encoding approach, the Quantum Lattice model, and FRQI.

\subsection{Quantum Computing}\label{subsec:quantum}

Quantum computing represents a revolutionary approach to information processing that harnesses the unique properties of quantum mechanics. Unlike classical computers, which store information as binary bits that are 100\% deterministic (0 or 1), quantum computers use quantum bits or “qubits” that can exist in superposition—a combination of 0 and 1 simultaneously.

The state of a qubit is mathematically defined as a linear combination of two standard basis states in Hilbert space, $|0\rangle$ and $|1\rangle$ (which are themselves made of complex numbers), as:
\begin{align}
    |\psi\rangle = \alpha|0\rangle + \beta|1\rangle
    \label{eq:qstate_1}
\end{align}
where $\alpha$ and $\beta$ are the probability amplitudes of the basis states and are both complex numbers.
A linear combination of the state vectors $\ket{0}$ and $\ket{1}$ represents a state of \emph{superpostion}, which intuitively corresponds to the idea that the qubit may be in some combination of states $\ket{0}$ and $\ket{1}$ at the same time.  

When we measure a quantum state $\ket{\psi}$, the probability $P_i$ of observing it in state $\ket{\psi_i}$ is given by the square of the magnitude (modulus squared) of the amplitude $a_i$:
\begin{align}
P_i = \lvert a_i \rvert^2
\end{align}
with a constraint from 
Born's Rule~\cite{Hidary:2019} that states the sum of all squared amplitudes must sum to one:
\begin{align}
    \sum_{i=0}^{n-1}\lvert a_i \rvert^2 = 1
\label{eq:BornsRule}
\end{align}

\begin{figure}[h!]
    \centering
    \includegraphics[width=0.5\linewidth]{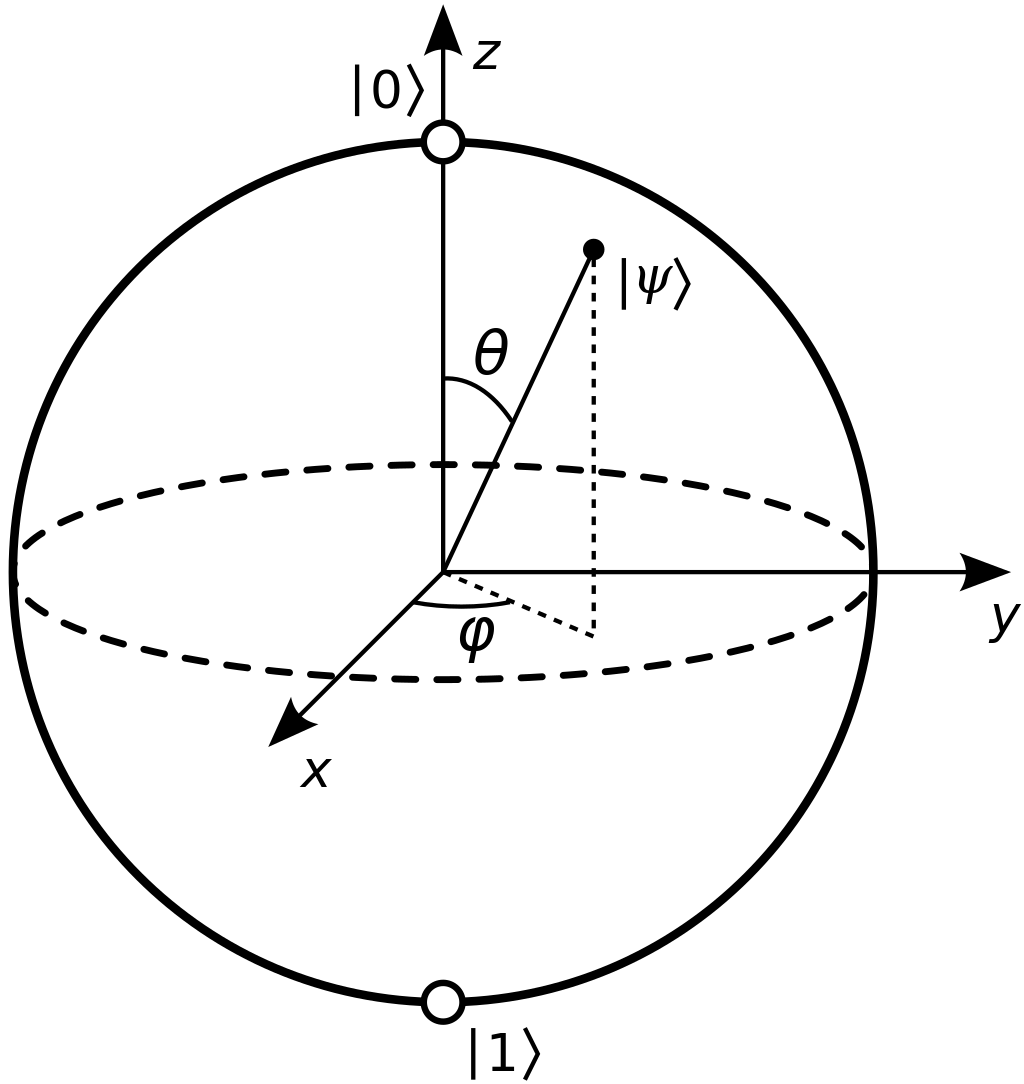}
\caption{The Bloch sphere is a geometric representation of a single qubit's quantum state. Any combination of $\alpha, \beta$ values in Eq.~\ref{eq:qstate_1} that conform to Born's Rule (Eq.~\ref{eq:BornsRule}) map a given quantum state $\ket{\psi}$ to a point on the surface of the sphere. \textit{Image source: Wikimedia Commons.} }
    \label{fig:bloch}
\end{figure}

A common way of visualizing the quantum state $\ket{\psi}$ is with diagram known as a Bloch Sphere~\cite{Bloch:1946}
as shown in Fig.~\ref{fig:bloch}. 
It is a geometric representation of a single qubit's quantum state shown as a point on a unit sphere.
Each quantum state $\ket{\psi}$ can be expressed in terms of two parameters: the azimuthal angle $\phi$ and the polar angle $\theta$, which correspond to the coefficients of the basis states $\ket{0}$ and $\ket{1}$ of a qubit's superposition.
The north and south poles represent the classical states $\ket{0}$ and $\ket{1}$, respectively. Points elsewhere on the sphere's surface represent superpositions, such as $\left ( \ket{0} + \ket{1} \right ) / {\sqrt{2}}$.
See Bethel et al.~\cite{Bethel:CGA_Quantum:2023} for a survey of visualization techniques for the quantum state of single- and multi-qubit systems.




Present-day quantum hardware faces problems like quantum decoherence, where delicate quantum superposition of states in a quantum system breaks down due to interactions with the surrounding environment~\cite{quantum_decoherence}, the physically complex challenge of maintaining gate fidelity, readout errors that can affect the reliability of computation results, etc. Works of Leymann F. et al.~\cite{Leymann_nisq} describe the challenges of implementing gate-based quantum algorithms on current error-prone and resource-constrained Noisy Intermediate Scale Quantum (NISQ) devices~\cite{Preskill2018quantumcomputingin}. 


\subsection{Hybrid Classical-Quantum Paradigm}\label{subsec:hybrid}
The prevailing view is that quantum computers will complement, rather than replace, classical systems, forming part of a larger hybrid high-performance computing (HPC) strategy where both systems work together to tackle currently unsolvable problems~\cite{hybrid_substrates, forbes_hybrid_future, forbes_quantum_revolution}. Classical computers will continue to be essential for most computational tasks, while quantum systems will enhance capabilities in specific areas where they provide distinct advantages. Thus, hybrid classical-quantum platforms are emerging as a promising trend for the coming decade.

Research on hybrid quantum computing is expanding, exploring various ways to integrate quantum and classical computing. For example, Phillipson et al.~\cite{hybrid_Phillipson} categorize hybrid computing approaches from a software architecture perspective, including models like Variational Quantum Algorithms (VQAs)\cite{VQA}, which iterate between quantum and classical computations to optimize solutions. Suchara et al.~\cite{hybrid_Suchara} highlight how classical supercomputers can support intermediate-scale quantum processors, enabling them to solve larger, more complex problems reliably despite hardware limitations.

This hybrid approach leverages the strengths of both quantum and classical systems, enhancing overall computational power and opening new possibilities for fields like optimization, cryptography, and machine learning.

\subsection{Benchmarking}\label{subsec:benchmarking}

In the world of classical HPC computing, benchmarks like LINPACK and benchmarking metrics like Millions of Floating Point Operations per second (MFLOPs) have emerged as a ``standard yardstick" for measuring and comparing the performance of computational platforms~\cite{dongarra2003linpack}.
In the quantum world, a completely different set of performance measures have emerged that are oriented towards understanding the limits of reliability of different size and complexity quantum programs on different quantum computing devices. 

A metric known as \emph{quantum volume} (\qv{}) defines the maximum size circuit of equal width (number of qubits) and depth (operations on a qubit) that may be reliably run on any given platform~\cite{Cross:Volumetric:2019}. 
The \qv{} for a given platform is determined either empirically through observation by running suitable benchmark programs or numerically by running machine models that incorporate noise, topology, and other characteristics of the particular platform.
%
%
While useful, \qv{} is somewhat limited and doesn't reflect the complexities present in gate-based quantum programs. 

Donkers et al., define a multidimensional hardware-agnostic benchmarking suite QPack~\cite{qpack}. 
It works by running a scalable Quantum Approximate Optimization Algorithm (QAOA) and evaluates the runtime, accuracy, and scale of a quantum computer. 
The QUARK framework by Finžgar et al.~\cite{Quark}, takes a rather application-oriented approach to gathering metrics. 
The QED-C benchmark suite by Lubinski et al.~\cite{qedc} is a set of quantum algorithms that measure multiple performance measures and produce output like volumetric benchmarking plots, fidelity comparisons, execution time, etc. Those benchmarks, however, target full-fledged quantum algorithms like Bernstein-Vazirani Algorithm ~\cite{Bernstein-Vazirani-benchmarking} and Quantum Fourier Transform~\cite{QFT}. 
Wack et al.~\cite{CLOPS} introduced a new measure, Circuit Layer Operations per Second (CLOPS), which measures how many quantum circuits a Quantum Processing Unit (QPU) can run in a given amount of time. A set of interesting metrics introduced by Tomesh et al.~\cite{supermarq} include calculations of qubit communication, critical depth, entanglement ratio, etc, and take a step towards quantum program profiling.\par

Langione et al.~\cite{quantum_benchmarking_article} investigate additional factors to enable comparisons between quantum computing's potential and our current computing capabilities.
Employing quantum algorithms in a workflow includes running multiple measurements with multiple ``shots" per measurement and some classical computation for preparation and readout. We end up with additional computations, both classical and quantum, apart from the actual quantum algorithm, and hence, measuring and benchmarking the overheads of quantum encoding is not trivial. 

In the context of hybrid computing, the classical encoding and image processing methods would indeed outperform the current quantum encoding techniques. However, it is essential to quantify and compare the existing quantum encoding models to help understand the key factors like resource demands, speed, and algorithm accuracy.

\section{Data Encoding Models}\label{sec:dataEnc}

To make use of classical data in a quantum computer, the classical data must be \emph{encoded} into the quantum state
as part of the quantum program initialization.
This state preparation 
can be done using different embedding or encoding techniques that translate classical data into quantum states in Hilbert Space. These encoding techniques might end up increasing exponentially in terms of circuit depth and width as well as runtime, and in the worst case,  will directly impact the promised speed-up of a quantum algorithm over classical. The undeniable importance of encoding paradigms has led to substantial research in this area.

\subsection{Qubit Lattice Model}\label{subsec:QubitLattice}

The Qubit Lattice model by Venegas-Andraca and Bose~\cite{QubitLattice} is one of the oldest algorithms for encoding and decoding image data in quantum computers. They described a straightforward method of directly encoding pixel intensities to the amplitude of the quantum state of individual qubits. There is no use of any quantum features like superposition and entanglement, limiting its applicability to a small number of processing methods, but large circuit widths.
\par
\begin{figure}[h!]
  \centering
  \includegraphics[width=\linewidth]{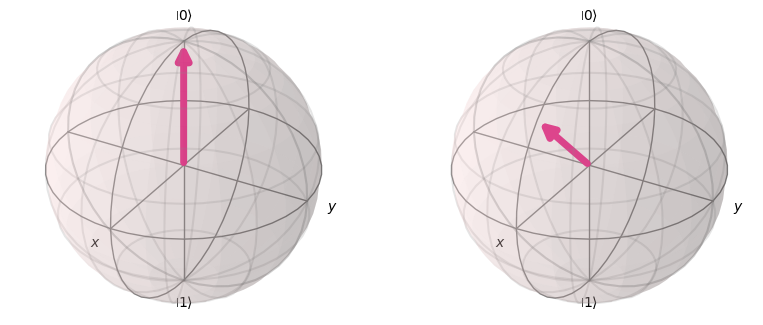}
  \caption{Visualizing the Qubit Lattice transformation using a single qubit to encode an angle of $\pi/4$ using a Bloch Sphere.}
  \label{fig:ql:bloch}
\end{figure}

The model has a rotation-based encoding approach, using the standard rotation gate $R_{y}$ (\ref{eq:RY}) to set individual pixel intensity to each qubit. 
\begin{align}
R_{y}(\theta) &= \begin{pmatrix}
                    \cos (\theta/2) & - \sin (\theta/2) \\
                    \sin (\theta/2) & \cos (\theta/2) \label{eq:RY}
                \end{pmatrix} & \text{(for an angle $\theta$)}
\end{align}
Figure \ref{fig:ql:bloch} shows this transformation on a pair of Bloch Spheres.  The first Bloch sphere shows the initial state of a qubit: $|0\rangle$. The next Bloch sphere represents the state of the qubit after applying the rotation gate by an angle $\theta$ (in this case $\pi/4$). 

Listing~\ref{listing:ql:encoder} generates a Qiskit circuit for the same logic. The initial step is to interpolate the pixel intensity from $[0, 255]$ to $[0, \pi]$ radians, which is used while applying the rotation gates.
\begin{lstlisting}[belowskip=\baselineskip, aboveskip=\baselineskip,caption={Qubit Lattice Encoder using single-qubit rotation gate around the Y axis.}, label={listing:ql:encoder}, name=QLEncoder, float=h, style=mystyle, language=Python]
from qiskit import QuantumCircuit
import numpy as np

def QL_encoder(qc:QuantumCircuit, angles:np.array):
    for i, ang in enumerate(angles):
        qc.ry(ang, i)
        
\end{lstlisting}

This approach encodes only the pixel intensities, not their position. Instead, the pixel position is implied by the ordering of qubits in a virtual lattice structure.
For example, in an image of size $N$ rows and $M$ columns, the first $M$ qubits hold the encoded pixel values for the first row of pixels, the second $M$ qubits hold the encoded values of the second row of pixels, and so forth. 
Layers of such a lattice structure can be used to encode RGB images, where each `layer’ holds the intensities of one color. Therefore, the circuit width grows to $n^{2}$ for an image of size $n \times n$. However, it has a considerably low circuit depth as it uses just one gate for encoding. 

Decoding is, however, a slightly complex process. For a pixel value $p$,
\begin{align}
p = 2 \times \arccos \left(\sqrt{P(j||Q_{p0}\rangle)}\right)
\end{align}\label{eq:ql:decode}
where,  
$P(j||Q_{p0}\rangle)$ is the conditional probability of observing a state $j$ given that the qubit $Q_{p0}$ (qubit associated with pixel $p$) is in state $|0\rangle$.






This is a purely classical operation with a complexity linear to the input size. The Qubit Lattice model laid the preliminary work in quantum image representation and paved the way for further research.
For an image of size $N$ by $M$ pixels, a total of $N \times M$ qubits are required for storing all pixel intensities. 
As a result, this form of encoding is useful only for very small images: current-day desktop-capable quantum simulators support $\mathcal{O}(30)$ qubits, and freely accessible platforms like those at IBMQ~\cite{IBMQ:2024} have $\mathcal{O}(100)$ qubits.

\subsection{Phase Encoding Model}\label{subsec:PhaseEncoding}

\begin{figure}[h!]
  \centering
  \includegraphics[width=\linewidth]{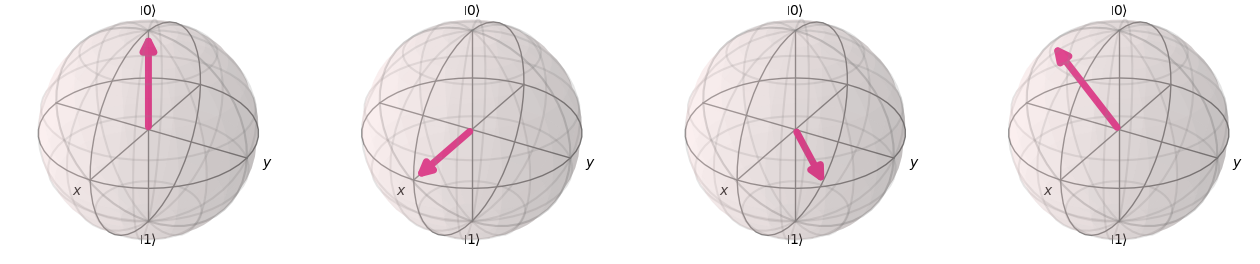}
  \caption{Visualizing the Phase Encoding technique using a single qubit to encode an angle of $\pi/4$ using a Bloch Sphere.}
  \label{fig:ph:bloch}
\end{figure}

The Phase Encoding technique also uses angles to encode the data points. These angles are related to the phase of the state's amplitude in the complex plane. This technique is found to be optimal for algorithms like Quantum Phase Estimation~\cite{nielsen_and_chuang} and Shor’s factoring~\cite{Shors}. To introduce a phase, the qubit is first brought into superposition by applying a Hadamard ($H$) gate (Eq.~\ref{eq:H}). At this stage, applying the $R_z$ rotation gate (Eq.~\ref{eq:RZ}) will rotate the qubit around the Z-axis with the appropriate angle, introducing phase. Applying a Hadamard again will put the qubit into the desired state, holding the value of classical data.
\begin{align}
\label{eq:H}
H &= \frac{1}{\sqrt{2}} \begin{pmatrix}
                            1 & 1 \\
                            1 & -1 
                        \end{pmatrix}
\end{align}
\begin{align}
R_{z}(\theta) &= \begin{pmatrix}
                    e^{-i(\theta/2)} & 0 \\
                    0                      & e^{i(\theta/2)} \label{eq:RZ}
                \end{pmatrix} & \text{(for an angle $\theta$)}
\end{align}
This full transformation $(H*R_z*H)$ results in the encoded state $\frac{1}{\sqrt{2}}(|0\rangle + e^{i\theta}|1\rangle)$.
This series of transformations are visually depicted in Fig.~\ref{fig:ph:bloch}.
The leftmost image shows the qubit in the initial $|0\rangle$ state. Then we apply a Hadamard gate and bringing the qubit into superposition (Eq.~\ref{eq:H}). Once the $R_z$ gate is applied (Eq.~\ref{eq:RZ}), the quantum state rotates around the Z-axis, which can be seen in the third image. The final, rightmost image shows the application of another Hadamard gate to see the phase effect as probability, which is our final state. Applying the last Hadamard gate helps to measure the phase as a probability. Reconstruction is similar to that of the Qubit Lattice model.

This Phase Encoding technique has the same limitations as the Qubit Lattice method, where one qubit per image pixel is required for storage.

\subsection{FRQI Model}\label{subsec:FRQI}

The Flexible Representation of Quantum Images
(FRQI) model by Le et al.~\cite{FRQI} encodes both pixel value and position as part of the quantum state. 
It encodes pixel values using amplitude encoding (\S\ref{subsec:QubitLattice}) but uses basis encoding (c.f.~\cite{Schuld:MLQC:2021}) to represent pixel coordinates. 


Using normalized superposition to store all the pixels, FRQI allows simultaneous operation on all the pixels and reduces the number of qubits used. The model encodes an image $I$ into the quantum state $\ket{I}$ as:
\begin{align}
\ket{I(\theta)} = \frac{1}{2^n}\sum_{i=0}^{2^{2n}-1} ( \cos\theta_i\ket{0} + \sin\theta_i\ket{1} ) \otimes \ket{i} \label{eq:FRQI}
\end{align}
here,
\begin{itemize}
    \item $\theta_i$ is the pixel value, typically $[0..255]$, normalized to the range $\left [ 0, \pi \right ]$
    \item $\ket{i}$ represents the pixel position
\end{itemize}




The encoding process consists of two primary steps: Initializing the state into a uniform superposition of all pixel positions and then applying a controlled rotation operation based on the pixel intensity value with $\theta$ defining the grayscale rotation.  The following is adapted from~\cite{IBM-FRQI}. 

The first step in FRQI encoding is to initialize the qubits to $|0\rangle^{\otimes2n+1}$.
Next is to put the state into superposition, except for the last qubit which is used to encode pixel intensity:
\begin{align}
    \ket{H} = \frac{1}{2^n}\ket{0} \otimes \sum_{i=0}^{2^{2n}-1}\ket{i}
\end{align}

Next, the controlled rotations are defined by: 
\begin{align}
R_i = \left ( I \otimes \sum_{j=0, j \neq i}^{2^{2n}-1}\ket{j}\bra{j} \right ) + R_y (2\theta_i) \otimes \ket{i}\bra{i}
\end{align}
where $R_y$ is the standard rotation matrices (Eq.~\ref{eq:RY}).




There are different ways of decomposing this complex transformation into single-qubit and standard rotation gates using $C^{2n}(R_{y}(2\theta_{i}))$ (multi-controlled rotation gate), one of which is presented in Listing ~\ref{listing:FRQIEncoder}. Every pixel value is interpolated into the $[0, \pi/2]$ to be used in the rotation phase. Adding an \texttt{MCRY} gate using 2 control qubits $(q_{c_0}, q_{c_1})$ and 1 target qubit $(q_t)$ from Qiskit produces the circuit,
\\

\begin{adjustbox}{width=\linewidth}   
\begin{quantikz}[]
\lstick{$q_t$}  & \gate{MCRY(\theta)} & \qw \\
\lstick{$q_{c_0}$} & \ctrl{-1}          & \qw  \\
\lstick{$q_{c_1}$} & \ctrl{-1}          &\qw 
\end{quantikz} = \begin{quantikz}[]
\qw  & \gate{R_y(\theta)} & \gate{X}  & \gate{R_y(-\theta)} & \gate{X}  & \qw \\
\qw & \qw                  & \ctrl{-1} & \qw                   & \ctrl{-1} & \qw  \\
\qw & \qw                  & \ctrl{-1} & \qw                   & \ctrl{-1} & \qw 
\end{quantikz}\label{circuit:mcry}
\end{adjustbox}
\break

This controlled rotation operation decomposed into a combination of $R_y(\theta)$ and \texttt{Toffoli (CCNOT)} gates, has to be applied incrementally to every pixel position. This is implemented by adding \texttt{CNOT} gates appropriately so that only the concerned qubits control the rotation.

\begin{lstlisting}[belowskip=-1.5\baselineskip, aboveskip=-1.5\baselineskip, caption={FRQI - Encoder logic}, label={listing:FRQIEncoder}, name=FRQIEncoder, float=h, style=mystyle, language=Python]
# inputs: array of angles based on interpolation
def FRQIencoder(angles:np.array):
    # Step 1
    qc = qiskit.QuantumCircuit()
    qc.id(q)    # 'q' Qregister = gray value
    qc.h(Q)     # 'Q' Qregister = coords

    controls_ = []
    for i in Q:
      controls_.append(i)
    
    # Step 2
    coords = np.ceil(math.log(len(angles), 2))

    for i, theta in enumerate(angles):
      index_bin = "{0:b}".format(i).zfill(coords)

      # enable appropriate control_qubits
      for k, qu_ind in enumerate(index_bin):
        if int(qu_ind):
          qc.x(Q[k])
                
      qc.barrier()

      qc.mcry(theta=2*theta, 
              q_controls=controls_, 
              q_target=q[0])

      qc.barrier()

      # disable control_qubits
      for k, qub_ind in enumerate(index_bin):
        if int(qub_ind): 
          qc.x(Q[k])
\end{lstlisting}
Recovering the original image is relatively more straightforward since FRQI states only contain real coefficients. \newline  
The following formula is used to decode the value of each qubit back from the FRQI state:
\begin{align}
\upsilon = \arccos \left(\sqrt{\frac{P(j||0\rangle)}{P(j||0\rangle) + P(j||1\rangle)}}\right) \label{eq:frqi:decode}
\end{align}
where,  
$P(j||0\rangle)$ is the conditional probability of observing a state $j$ given that the gray value of the qubit is in state $|0\rangle$.
Similarly, $P(j||1\rangle)$ is the probability of observing a state $j$ given that the gray value of the qubit is in state $|1\rangle$.
FRQI performs better than the Qubit Lattice and Phase Encoding models as it can process all pixels of the image simultaneously.

\vspace{25mm}

There are, however, some drawbacks:
\begin{itemize}
    \item Using a single qubit for encoding color information makes it challenging to design algorithms that require explicit color data, such as partial color operations and statistical color operations ~\cite{QIR_review, NEQR}.
    \item The intensity-to-amplitude representation is probabilistic in nature, and hence, it cannot be accurately measured in finite measurements.
\end{itemize}
FRQI is one of the promising encoding methods and there are several models built on top of it, like works of Zhang et al. - Novel Enhanced Quantum Representation (NEQR)~\cite{NEQR}, Multi-Channel Representation for Quantum Image (MCRQI) by Sun et al.~\cite{MCRQI}, and QPIXL by Amankwah et al.~\cite{QPIXL}

\subsection{Other Encoding Methods and Our Study}
\label{sec:other_encoding}

The three encoding methods discussed earlier use angle encoding to represent floating-point pixel values as phase or amplitude in the quantum state, with FRQI adding basis encoding for pixel position.

Numerous approaches exist for encoding classical data on quantum platforms. For example, Yan and Venegas-Andraca~\cite{Yan-Venegas:TwentyYears:2024} summarize over 20 encoding methods applied to image data, while Schuld and Petruccione~\cite{Schuld:MLQC:2021} explore encoding within quantum machine learning. Some algorithms for quantum linear algebra, such as the HHL algorithm~\cite{harrow2009quantum}, utilize amplitude encoding. Across these methods, angle encoding—using amplitude or phase—is a common technique for representing continuous data in $\mathbb{R}^n$. 

In contrast, basis encoding represents and manipulates the probabilities of computational basis states. Image encoding methods like NEQR use basis encoding to map real-valued pixel data, with the pixel resolution defined during state preparation. For instance, 8-bit pixel values in
$[0\dots255]$ require 8 qubits, allowing a tensor product of pixel positions and values to produce the final image encoding $\ket{I}$~\cite{NEQR}.

Our study focuses on methods that use angle encoding to represent real-valued pixel data, aligning with other real-valued data encodings and avoiding design complexities like those in NEQR. The methods in this study vary in complexity, affecting quantum state preparation, processing, and the approach to measurement, readout, and decoding back to classical data.

\section{Evaluation Objectives and Methodology}\label{sec:evaluation}
\subsection{Computational Software and Environment}
Primary development and initial tests were done on commodity hardware with Intel's six-core i7-11800H processor with a 2.3 GHz clock speed, supported with 16GB memory and 18MB of on-chip cache. Further extensive studies were carried out on the Perlmutter Cray EX supercomputer at the NERSC facility.\footnote{\url{https://docs.nersc.gov/systems/perlmutter/architecture/}} Quantum resources at IBM were used to gather results on an actual QPU for some curated test cases:
\begin{itemize}
    \item Experiment- FRQI for input sizes $[4, 16, 64, 256]$
    \item No. of shots- 10,000
    \item Machine- IBM Osaka (5K CLOPS, 2.8\% EPLG)
\end{itemize}
The project is developed in Python \textit{(version 3.11)} using IBM Quantum's Qiskit SDK ~\cite{Qiskit} \textit{(version 1.0.2)}. The code is publicly available\footnote {\url{https://github.com/simplysudhanshu/bits_to_qubits}}.

\subsection{Metrics}\label{subsec:metrics}

\begin{enumerate}
    \item Encoding Runtime:
        \newline
        Runtime is one of the most fundamental and vital ways to compare algorithms and approaches. It is directly related to efficiency and plays an important role in comparing classical and quantum approaches. The benchmarking framework collects the processing time required for encoding, which covers preparing the interpolated angles from the input vector and generating the state-preparation circuit.
\\
    \item Correctness Assessment:
    \newline
        Two different metrics, Accuracy and Precision, are used to understand the algorithm's correctness. Precision gives the number of correctly reconstructed values:
        \begin{align*}
            P = \frac{\text{Number of pixels with expected reconstruction}}{\text{Total number of pixels}}
        \end{align*}
        On the other hand, accuracy is a term associated with the deviation of the reconstructed value from the inverted pixel intensity. We calculate the error value $(E)$, which is $1-Accuracy$, based on the observed and expected value $(V)$ of each pixel,
        \begin{align*}
            E = \frac{|V_{observed} - V_{expected}|}{V_{expected}}
        \end{align*}
    \item Circuit Characteristics:
    \newline
    The standard metrics for evaluating circuits implemented using different encoding methods are circuit depth and circuit width. Circuit depth refers to the number of layers of operations (gates), while circuit width indicates the number of qubits used.
\\
    \item Circuit Fidelity:
    \newline
        Quantum fidelity measures the ``closeness" or similarity of the quantum states, typically based on their density operators. For quantum states $\rho$ and $\sigma$, Fidelity ($F$):
        \begin{align*}
            F_{pure} = |\langle\psi_{\rho}|\psi_{\sigma}\rangle|^{2}
        \end{align*}
        Qiskit implements Hellinger fidelity between two counts distributions, defined as $(1-H^{2})^{2}$ where $H$ is the Hellinger distance~\cite{qiskit_hellinger_fidelity}.
\\
    \item FRQI Probabilistic Experiment:
    \newline
        FRQI is highly probabilistic in nature, meaning it is difficult to accurately recover the image in a finite number of shots. This metric aims to capture the 
        correlation between the number of shots and accuracy. Increasing the number of shots improves the accuracy but also directly impacts the algorithm runtime. In this experiment, we measure the correlation between these three entities.
\\
    \item SupermarQ features~\cite{supermarq}:
    \begin{itemize}[nosep,leftmargin=0pt,labelindent=0pt]
        \item \textsc{Communication}
        \newline
            The degree of communication between qubits is an important metric in a quantum circuit, especially when the circuit is executed on actual quantum hardware and the underlying physical architecture affects the compilation. Based on qubit-interaction graphs, the communication is calculated as,
            \begin{align*}
                C = \frac{\sum^{N}_{i}d(q_{i})}{N(N-1)}
            \end{align*}
            for a circuit with $N$ qubits and $d(q_{i})$ is the normalized average degree of qubit calculated by two-qubit operations in the circuit.
            
        \medskip
        \item \textsc{Critical Depth}
        \newline
            The two-qubit gate operations in the circuit directly influence the runtime due to their higher execution time and accuracy because of the higher error rates of a quantum circuit. This metric relates to the number of such gates on the longest path of the circuit and is calculated as,
            \begin{align*}
                D = n_l/n
            \end{align*}
            where $n_l$ is the number of two-qubit interactions on the longest path, and $n$ is the total number of such interactions.
        
        \medskip
        \item \textsc{Entanglement Ratio}
        \newline
        While the previous measures focus on single-qubit gates, achieving scalability in quantum computing requires a high level of quantum entanglement. The `amount' of entanglement in a circuit is important for demonstrating quantum advantage, though it can be challenging to measure accurately. 
        A simple measure of entanglement is the ratio of two-qubit interactions $(n)$ to the total number of gate operations in the circuit ($n_t$):
        \begin{align*}
                E = n/n_t
            \end{align*}
            This ratio $E$ gives the percentage of two-qubit operations relative to all gate operations in the circuit.

        \medskip
        \item \textsc{Parallelism}
        \newline
        Parallelism helps to measure how many gates can be executed simultaneously in a circuit. While increased parallel operations can improve efficiency, it may also introduce an error called ``crosstalk," which can reduce performance \cite{Zhao:Crosstalk:PRXQuantum:2022}. Parallelism for a circuit with $N$ qubits, $d$ depth, and $n_t$ total gate operations as:
            \begin{align*}
                P = \left(\frac{n_t}{d} - 1\right) \frac{1}{N-1}
            \end{align*}
        

        \medskip
        \item \textsc{Liveness}
        \newline
        To track a qubit's activity throughout the circuit, we define a liveness matrix that indicates, for each layer, whether a qubit is involved in an operation. Liveness is calculated as
         \begin{align*}
                L = \frac{\sum_{ij}A_{ij}}{nd}
            \end{align*}
             where $A$ is the liveness matrix of a $N$-qubit circuit of depth $d$.
             Each entry $A_{ij}=1$ if quibit $i$ is active in layer $j$, and 0 otehrwise. Liveness reflects how often each qubit is engaged across the circuit's layers. 
             
    \end{itemize}
\medskip
    \item Backend comparisons:
    \newline
        For a comparative analysis, some key metrics, like runtime and accuracy, are measured on different backends, such as a pure simulator, a noisy simulator, and IBMQ hardware.
        
\end{enumerate}

\subsection{Methodology}\label{subsec:methodology}
Input data for our experiments is 
a randomized $n \times n$ grayscale image where each pixel has value $\in [0, 255]$. The input image sizes are chosen based on the algorithm's input constraints (FRQI can accurately encode square images-$2^{n} \times 2^{n}$ only) and available computation resources (memory requirements for classical simulations of quantum circuits),
\begin{align*}
    n \in 
    \begin{cases}
        [2, 3, 4, 5] & \text{Qubit Lattice and Phase Encoding} \\
        [2, 4, 8, 16] & \text{FRQI Model}
    \end{cases}
\end{align*}

Once encoded, we implement a simple unary operation on the pixel data: invert the pixel value.
The objective of this simple operation is consistent with our choice of encoding methods (see \S\ref{sec:other_encoding}) to focus on a limited set of operations on real-valued data so as to focus most of the performance measurement on the encoding/decoding portions of the method. 
\section{Findings and Discussion}\label{sec:findings}

This section presents a comparative analysis of various encoding techniques based on key performance metrics. While most results are obtained from simulations, some are derived from experiments on IBM hardware for backend comparisons. Special emphasis is placed on the FRQI encoding model, with detailed studies focusing on its probabilistic experiments.

\subsection{Encoding Runtime}
\label{sec:results:runtime}

\begin{figure}[b]
  \centering
  \includegraphics[width=\linewidth, height=0.75\linewidth]{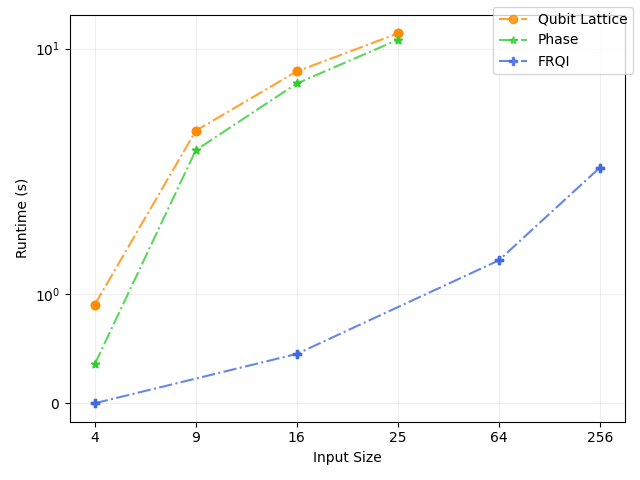}
  \caption{Encoding runtime analysis for the whole range of input sizes and all three encoding techniques.}
  \label{fig:comp:runtime}
\end{figure}

We begin with a fundamental performance measure: runtime. Here, we measure the time required to build a quantum circuit that encodes the input data. Figure \ref{fig:comp:runtime} shows a comparison of encoding runtimes. Angle-based encoding methods exhibit similar trends, while methods with lower qubit requirements, such as FRQI, demonstrate faster runtimes. This efficient performance can lead to better resource utilization, making these methods more practical for hybrid classical-quantum computing applications.

Another important factor is the time needed to transpile these circuits for specific backends. Transpilation optimizes a quantum circuit for a particular backend by considering its topology, native operations, connectivity, and circuit depth. Although optimizations are possible, transpilation can still take a significant amount of time.


\subsection{Circuit Characteristics}\label{sec:results:width_depth}
\begin{figure}[h!]
\begin{subfigure}{\linewidth}
  \centering
  \includegraphics[height=0.65\linewidth, width=\linewidth]{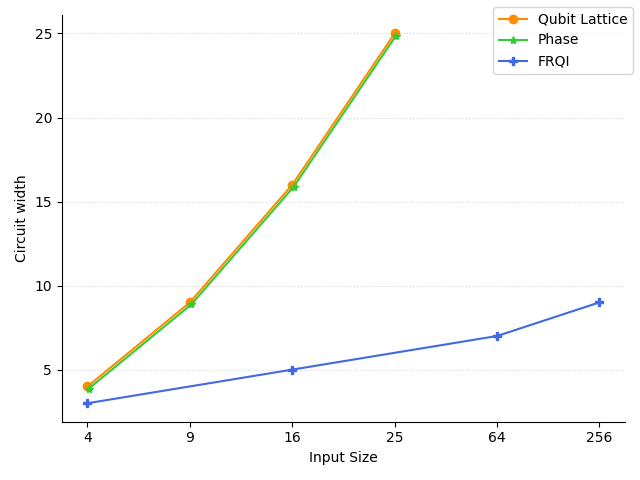}
  \caption{Width (No. of qubits)}
  \label{fig:comp:width}
\end{subfigure}
\begin{subfigure}{\linewidth}
  \centering
  \includegraphics[height=0.65\linewidth, width=\linewidth]{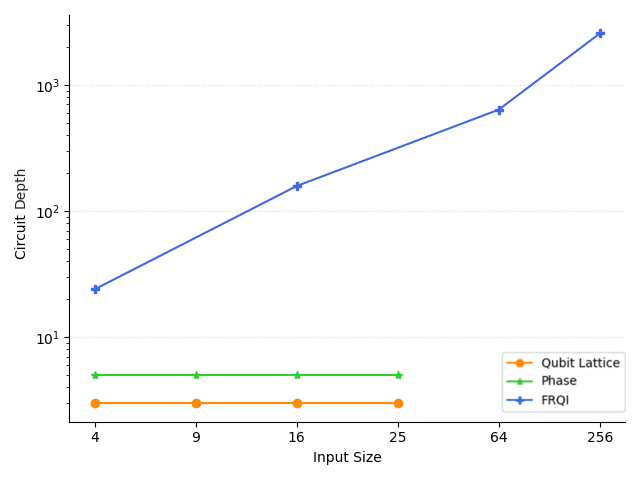}
  \caption{Depth (No. of operations)}
  \label{fig:comp:depth}
\end{subfigure}
\caption{Understanding the trend of the core properties of a quantum circuit as we scale the input size.}
\label{fig:comp:circuit}
\end{figure}

The fundamental characteristics of a quantum circuit are its depth and width. Comparing these metrics helps to understand the resource demands as input sizes grow. These values can be easily calculated using Qiskit's circuit representation. The depth of rotation-based encodings remains constant across experiments, highlighting their advantage (see Figure~\ref{fig:comp:depth}). In terms of width, FRQI outperforms the other encodings (see Figure~\ref{fig:comp:width}), requiring fewer qubits. This lower qubit requirement reduces hardware demands, control complexity, and potentially improves computation fidelity.

\subsection{Correctness Assessment}
\label{sec:results:accuracy}

\begin{table*}
  \centering
\begin{tabular}{@{}c|ccc|ccc@{}}
\toprule
  \multirow{2}{*}{\textbf{Input Size}} &
  \multicolumn{3}{c|}{\textbf{Precision}} &
  \multicolumn{3}{c}{\textbf{Mean Error Value}} \\ \cmidrule(l){2-7} 
 &
  \multicolumn{1}{p{1.9cm}|}{\textbf{Qubit Lattice}} &
  \multicolumn{1}{p{2cm}|}{\textbf{Phase Encoding}} &
  \multicolumn{1}{p{1.8cm}|}{\centering\textbf{FRQI}} &
  \multicolumn{1}{p{1.9cm}|}{\textbf{Qubit Lattice}} &
  \multicolumn{1}{p{2cm}|}{\textbf{Phase Encoding}} &
  \multicolumn{1}{p{1.8cm}}{\centering\textbf{FRQI}} \\ \midrule
2x2   & 100.00 \% & 100.00 \% & 75.00 \% & 0.00 & 0.00 & 0.25 \\[0.3cm]
3x3   & 77.00 \%  & 66.26 \%  & x        & 0.22 & 0.33 & x    \\[0.3cm]
4x4   & 43.75 \%  & 68.75 \%  & 31.25 \% & 0.56 & 0.31 & 0.94 \\[0.3cm]
5x5   & 56.00 \%  & 44.00 \%  & x        & 0.44 & 0.56 & x    \\[0.3cm]
8x8   & x         & x         & 22.93 \% & x    & x    & 2.38 \\[0.3cm]
16x16 & x         & x         & 20.46 \% & x    & x    & 4.34 \\ \bottomrule
\end{tabular}
\caption{Correctness assessment of all three encoding techniques across the whole range of problem sizes.}
\label{tab:comp:accuracy}
\end{table*}
Precision refers to the number of correctly mapped (and reconstructed) data points, while the mean error value indicates how much the results deviate from expected values. Table \ref{tab:comp:accuracy} provides a detailed comparison. A lower mean error indicates higher accuracy, with reconstructed values closely matching the original values. The encoding methods show minimal errors, and increasing the number of shots further improves accuracy.

\subsection{FRQI probabilistic experiment}

\begin{figure}[!h]
  \centering
  \includegraphics[width=\linewidth]{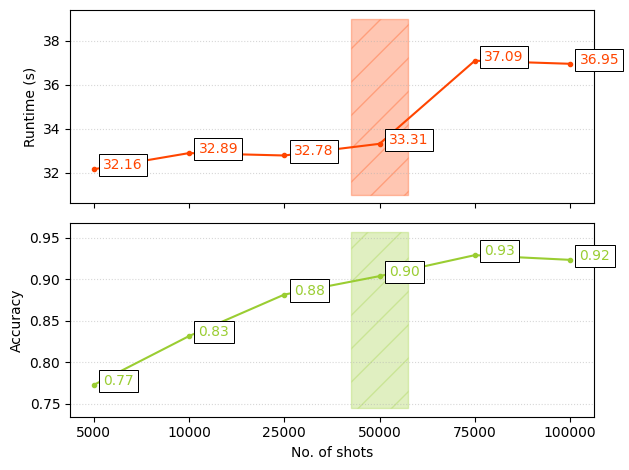}
  \caption{FRQI circuit for $16\times16$ image. The shaded area indicates optimal accuracy-runtime efficiency trade-off.}
  \label{fig:frqi:shots}
\end{figure}

As described in Section~\ref{subsec:metrics}, it's useful to examine how the number of measurements, or 'shots,' affects FRQI performance. Figure~\ref{fig:frqi:shots} shows FRQI’s runtime and accuracy as shot count increases. This analysis uses the highest configuration of the FRQI experiment suite: a $16 \times 16$ image with 9 qubits and a depth of 2554. The runtime shown represents the total algorithm runtime, as encoding time remains constant but simulation time varies. The plot reveals that increasing the number of shots enhances accuracy by improving precision in probability calculations, thus reducing error.
The area of interest shaded in this graph shows the most efficient performance using a ratio of \textit{accuracy / runtime}. This indicates the configuration that gives the most ``bang for the buck," where the trade-off between accuracy and runtime is most efficient.

\subsection{Fidelity comparisons}
\begin{figure}[!h]
  \centering
  \includegraphics[width=\linewidth]{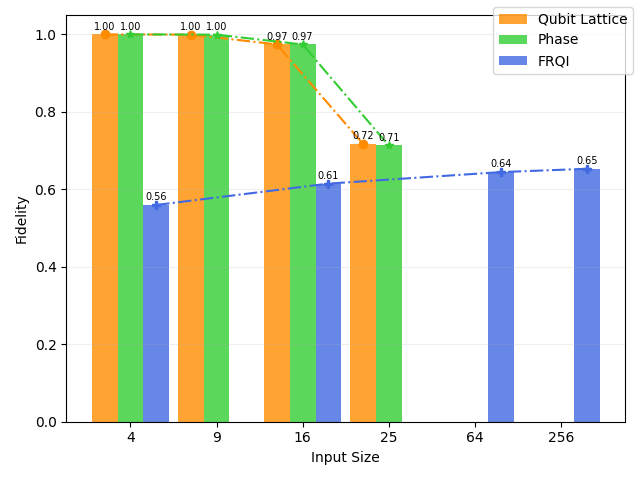}
  \caption{Comparing circuit fidelity of all three encoding techniques across a whole range of problem sizes at 5,000 shots.}
  \label{fig:comp:fidelity}
\end{figure}

To examine how key properties of a quantum circuit scale with input size, we use Hellinger fidelity~\cite{qiskit_hellinger_fidelity} as a metric to assess how closely the circuit approximates target quantum states. This metric compares probability distributions by calculating the fidelity between the ideal state vector and the quasi-probabilities obtained experimentally, allowing insights into the effectiveness of each encoding technique used. In Figure \ref{fig:comp:fidelity}, fidelity values for the three encoding methods—FRQI, Qubit Lattice, and Phase Encoding—are illustrated across varying input sizes. FRQI shows consistently lower fidelity, particularly at smaller input sizes, possibly due to the limited number of shots, which may reduce accuracy in probability estimation. Conversely, the Qubit Lattice and Phase Encoding models demonstrate high fidelity across input sizes, although both experience a sudden drop at the largest input size. This drop is likely due to the exponential growth in states within the circuit’s state vector, which poses scaling challenges and may limit the fidelity achievable by these encoding methods at larger scales.

\begin{figure*}[!t]
\begin{subfigure}{.33\textwidth}
  \centering
  \includegraphics[width=\linewidth, height=0.95\linewidth]{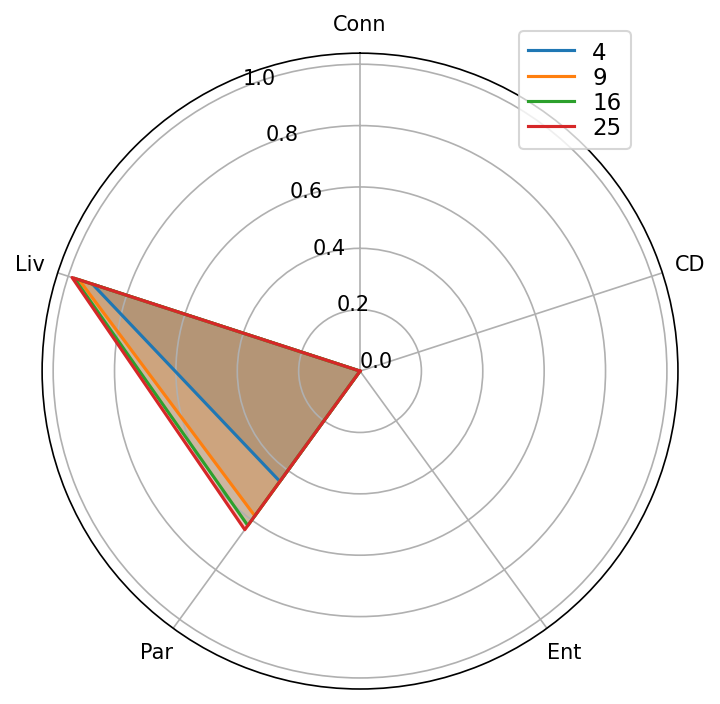}
  \caption{Qubit Lattice}
  \label{fig:comp:supermarq:ql}
\end{subfigure}%
\begin{subfigure}{.33\textwidth}
  \centering
  \includegraphics[width=\linewidth, height=0.95\linewidth]{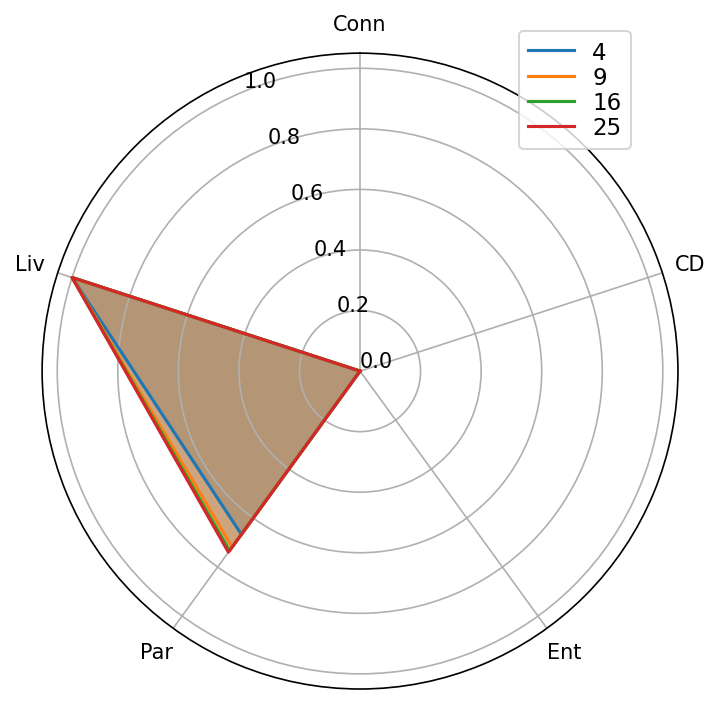}
  \caption{Phase Encoding}
  \label{fig:comp:supermarq:ph}
\end{subfigure}%
\begin{subfigure}{.33\textwidth}
  \centering
  \includegraphics[width=\linewidth, height=0.95\linewidth]{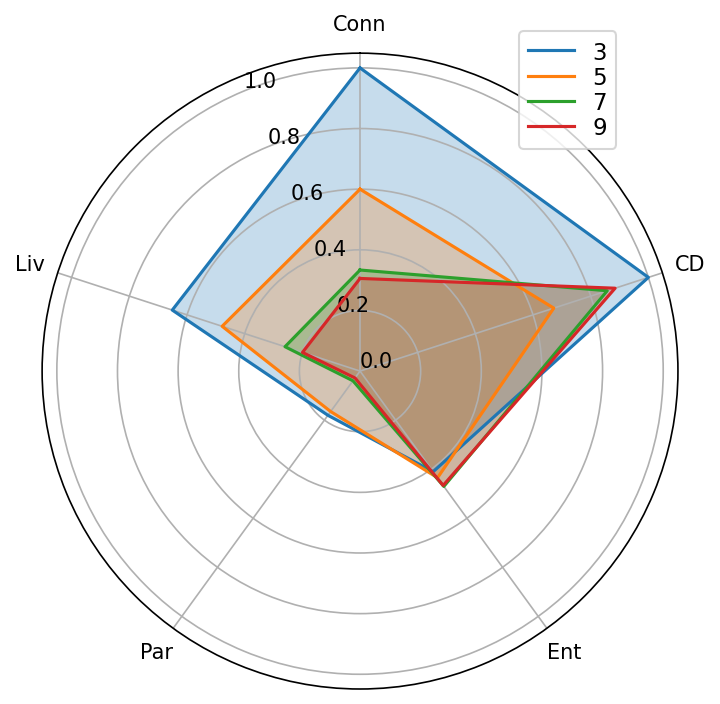}
  \caption{FRQI}
  \label{fig:comp:supermarq:frqi}
\end{subfigure}
\caption{SupermarQ features for all three encoding experiments. Detailed legend: Conn - \textit{Communication}, CD - \textit{Critical Depth}, Ent - \textit{Entanglement Ratio}, Par - \textit{Parallelism}, Liv - \textit{Liveness}}
\label{fig:comp:supermarq}
\vspace{-5mm}
\end{figure*}

\subsection{SupermarQ}
Figure~\ref{fig:comp:supermarq} 
illustrates how circuit configuration impacts noise. The Qubit Lattice and Phase Encoding models have higher liveness due to their shorter circuit lengths and consistent use of all qubits in each layer. Phase Encoding also exhibits high parallelism by packing many operations into a small circuit depth, which introduces noise events like `cross-talk' that can disrupt the quantum state~\cite{cross-talk}. In FRQI, high connectivity, critical depth, and entanglement result from its large circuit and use of \texttt{CCX} gates, contributing to accuracy drops as multi-qubit gates are more prone to noise. FRQI has low parallelism due to numerous \texttt{CNOT} gates, even after decomposition, and lower liveness, as controlled operations often only involve a subset of qubits, leaving others idle.

\subsection{Backend Comparison}
\begin{figure}[b!]
\vspace{-8mm}
\begin{subfigure}{\linewidth}
  \centering
  \includegraphics[width=\linewidth]{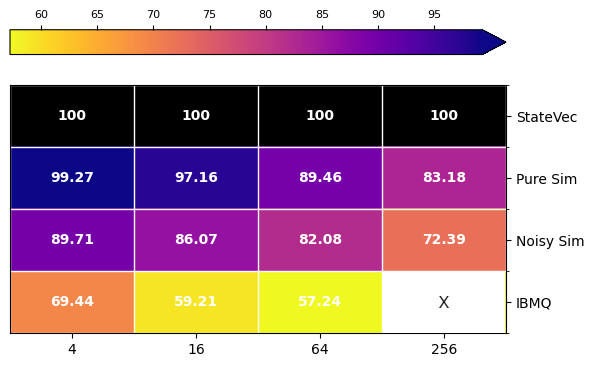}
  \caption{Accuracy (\%)}
  \label{fig:backend:accuracy}
\end{subfigure}
\begin{subfigure}{\linewidth}
  \centering
  \includegraphics[width=\linewidth]{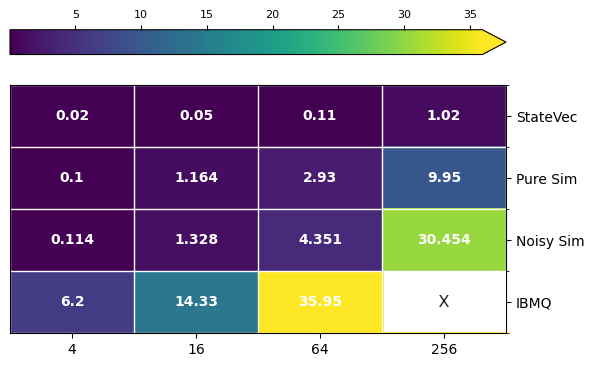}
  \caption{Runtime (s)}
  \label{fig:backend:runtime}
\end{subfigure}
\caption{FRQI model executed on multiple hardware for all input sizes for 10,000 shots.}
\label{fig:backend}
\end{figure}

Results from real quantum machines validate simulation accuracy and highlight hardware-specific limitations. We extended our investigation of the FRQI encoding model by running experiments on various backends, including quantum simulators and actual quantum hardware hosted on the IBM Quantum Platform. For the FRQI model, two key metrics—accuracy and runtime—were evaluated at all input sizes. Figure \ref{fig:backend:accuracy} shows accuracy results, while Figure \ref{fig:backend:runtime} displays total execution times.

The `StateVec' backend is a classical computation based on the circuit’s state vector, calculated without simulating shots. The Pure and Noisy simulators are Qiskit AerSimulators running locally, providing comparable accuracy and runtime. In contrast, IBM’s actual quantum machine shows lower accuracy due to real-world decoherence and gate errors.

At the maximum input size of $16 \times 16$, the circuit depth reaches 2554 before transpilation due to sequential controlled rotations. However, after targeting the IBMQ backend, the transpiled circuit’s payload exceeded the hardware's execution capacity, even at maximum optimization, resulting in an `x' mark indicating an unsuccessful run.

\section{Conclusion}\label{sec:conclusion}
This study examined the performance and characteristics of three quantum encoding techniques: FRQI, Qubit Lattice, and Phase Encoding, focusing on computational cost, runtime, and accuracy. Key findings include:

\begin{itemize} 

\item \textbf{Accuracy}: For lower shot counts, Qubit Lattice and Phase Encoding provide higher accuracy than FRQI. However, with a higher shot count, FRQI can achieve comparable or superior accuracy with lower circuit width but greater circuit depth.

\item \textbf{Runtime}: FRQI is the fastest among the three techniques in both encoding and overall runtime due to its lower qubit requirements, as it uses a single qubit per pixel. Circuit width and depth strongly influence runtime across all methods.

\item \textbf{Circuit complexity}: Circuit complexity varies across techniques. FRQI has low circuit width but significantly higher depth, which increases noise. The SupermarQ plots provide further insight into noise resilience based on circuit structure.

No encoding technique universally excels across all metrics. The choice of encoding method depends on the specific processing algorithm and requires careful consideration of circuit dimensions, shot count, and computational resources.

While overhead is inherent to these techniques, it can be managed. For maximum accuracy, Qubit Lattice and Phase Encoding are effective if there are no time or hardware constraints. However, in the current NISQ era, where quantum hardware is limited and costly, FRQI offers a practical balance of accuracy and minimal qubit requirements, making it suitable for near-term quantum image processing applications.

This study contributes to the understanding of quantum encoding techniques, supporting informed choices and future advancements in quantum computing. Future work could apply these encoding methods in complex quantum image processing algorithms to further assess the impacts on circuit complexity, runtime, and accuracy.
\end{itemize}

\section*{Acknowledgment}

This research used resources of the National Energy Research Scientific Computing Center (NERSC), a Department of Energy Office of Science User Facility using NERSC award DDR-ERCAP0026767.





\printbibliography

\end{document}